\documentclass[aip,graphicx]{revtex4-1}

\usepackage{graphicx}
\usepackage{epstopdf}

\draft 

\begin{document}





\title{Self-similar transmission properties of aperiodic Cantor potentials in gapped graphene}

\author{R. Rodr\'iguez-Gonz\'alez}

\email[Corresponding author:]{ rodriguezglez.r@fisica.uaz.edu.mx}

\author{I. Rodr\'iguez-Vargas}

\affiliation{Unidad Acad\'emica de F\'isica, Universidad Aut\'onoma de Zacatecas, Calzada Solidaridad Esquina Con Paseo La Bufa S/N, 98060 Zacatecas, Zac., M\'exico.}

\author{D. S. D\'iaz-Guerrero}

\author{L. M. Gaggero-Sager}

\affiliation{Facultad de Ciencias, Universidad Aut\'onoma del Estado de Morelos, Av. Universidad, Col. Chamilpa 62209 Cuernavaca, Morelos, M\'exico.}

\date{\today}

\begin{abstract}
We investigate the transmission properties of quasiperiodic or aperiodic structures based on graphene arranged according to the Cantor sequence. In particular, we have found self-similar behaviour in the transmission spectra, and most importantly, we have calculated the scalability of the spectra. To do this, we implement and propose scaling rules for each one of the fundamental parameters: generation number, height of the barriers and length of the system. With this in mind we have been able to reproduce the reference transmission spectrum, applying the appropriate scaling rule, by means of the scaled transmission spectrum. These scaling rules are valid for both normal and oblique incidence, and as far as we can see the basic ingredients to obtain self-similar characteristics are: relativistic Dirac electrons, a self-similar structure and the non-conservation of the pseudo-spin. This constitutes a reduction of the number of conditions needed to observe self-similarity in graphene-based structures, see D\'iaz-Guerrero et al. [D. S. D\'iaz-Guerrero, L. M. Gaggero-Sager, I. Rodr\'iguez-Vargas, and G. G. Naumis, arXiv:1503.03412v1, 2015]. 
\end{abstract}


\maketitle 

\section{Introduction}
In the last decade, the aperiodic modulation has attracted great interest due to its peculiar electronic and optoelectronic properties \cite{EMacia2006RPP, EMacia2012RPP} such as the transmission properties that appear highly fragmented displaying self-similar patterns along with a fractal-like electronic spectra,\cite{RPAlvarez2001JPCM} different to what occurs in periodic superlattices. All these characteristics are a direct consequence of the quasiperiodicity that underlies in these aperiodic structures.

Quasiperiodic or aperiodic systems are structures that can be considered as fractals,\cite{JFeder1988} homogeneous and self-similar geometrical objects, that also are defined as structures ordered but without periodicity. In this context, the aperiodic structures or also commonly known as deterministic media are classified in two group according to the constructions algorithm. Those generated by means of a substitutional procedure named as substitutional sequences (Fibonacci, Thue-Morse, Double-Period, Rudin-Shapiro and Silver-Mean) and those generated in a similar way to fractal sets, \cite{JFeder1988} named as model fractal structures (Cantor sets and Koch fractals). Among the peculiar characteristics of the quasiperiodic structures excels the so-called self-similarity.\cite{RPAlvarez2001JPCM} This property means that one or more parts of a structure or object have exactly or approximate the same shape of the whole, and even more if these parts are magnified at different scales the result is the entire object.  

Self-similarity is found everywhere, from geographic structures, statistical natural processes to complex networks. \cite{JFeder1988} In terms of length scales, we can say that self-similarity is found in systems of the size of the atom to systems of the size of the universe.\cite{JFeder1988} Specifically, this outstanding property was recognized in the coasts, in the distribution of species, in the growth of trees, rivers and lungs, to mention a few.\cite{JFeder1988, JHarte1999Science, CEloy2011PRL} Within this context, the aforementioned self-similar parts of a structure can be reproduced at different scales, so it is natural to assume that there are scaling rules that relates the scalability of thereof. However, in most of the multiple reports that we can find in the literature about solid-state systems, self-similarity is let as a matter of visual perception, and rarely the scaling rules are reported.\cite{ALavrinenko2002PRE}

In this sense, since the discovery of graphene\cite{KNovoselov2004Science, KNovoselov2005Nature, YZhang2005Nature} researchers have begun to deal with aperiodic structures based on this prominent material.\cite{SSena2010JPCM, SMukhopadhyay2010PSSB, PZhao2011APL,AKorol2013PSS,WTLu2013PLA, TMa2012APL, ZZhang2012APL,YXu2013JPCM,HHuang2013JAP,CLi2013APL,HLiu2013JAP,AKorol2014LTP,LSun2010JPCM,
YZhang2012JPDAP,DLiu2014JPDAP,XChen2013JPDAP,YPZhang2014PLA,YPZhang2014CPB} For instance, the Fibonacci sequence is studied in monolayer\cite{SSena2010JPCM} and bilayer\cite{SMukhopadhyay2010PSSB} graphene superlattices. In particular, a striking self-similar behaviour is reported for different generations number of the aperiodic sequence, \cite{SSena2010JPCM} it is also found that the transmission spectra are fragmented, and even more appear in groups and the low-energy conductance is considerably amending.\cite{SMukhopadhyay2010PSSB} With respect to the Cantor set, a few works have been devoted in the same direction. \cite{LSun2010JPCM,YZhang2012JPDAP,DLiu2014JPDAP} Specifically, a magnetic modulation has been implemented to study the transport properties through graphene-based fractal barriers. \cite{LSun2010JPCM} Despite the particularities of these reports, we can see that they have a common aspect, a quasiperiodic structure with self-similar properties, which are manifested in the transmission spectra. However, the scaling rules of the self-similar patterns are not found, and once again self-similarity is considered as a matter of visual perception. Thus, a simple question arises, what are the appropriate scaling rules that can reproduce the self-similar patterns?.

Following the lines of a previous work,\cite{DSDan2015} we obtain the scaling rules of the self-similar transmission patterns of Cantor graphene structures (CGSs). In this case CGSs are constructed by simply scaling (spatially) the barriers of the system  according to the triadic Cantor set rules. Scaling rules between generations, lengths of the system and heights of the barriers are obtained. These scaling properties are valid for both normal and oblique incidence. We also carry out a comparative analysis of the scaling expressions in substrate-based CGSs (SCGSs), electrostatic CGSs (ECGSs) and Al$_{x}$Ga$_{1-x}$As/GaAs semiconductor structures (SSs), finding a better scaling in the case of SCGSs. Our results indicate that conditions like the non-symmetry (spatially speaking) and the scaling in the energy axis of the structure  are not necessary to obtain self-similarity in the transmittance. \cite{DSDan2015} Therefore, as far as we can see, the conditions to obtain self-similarity in the transmission properties are: relativistic Dirac electrons, a self-similar structure and the non-conservation of the pseudo-spin (electrons in SCGSs).  

\section{Methodology}
\begin{figure}[h]
\includegraphics[scale=0.28]{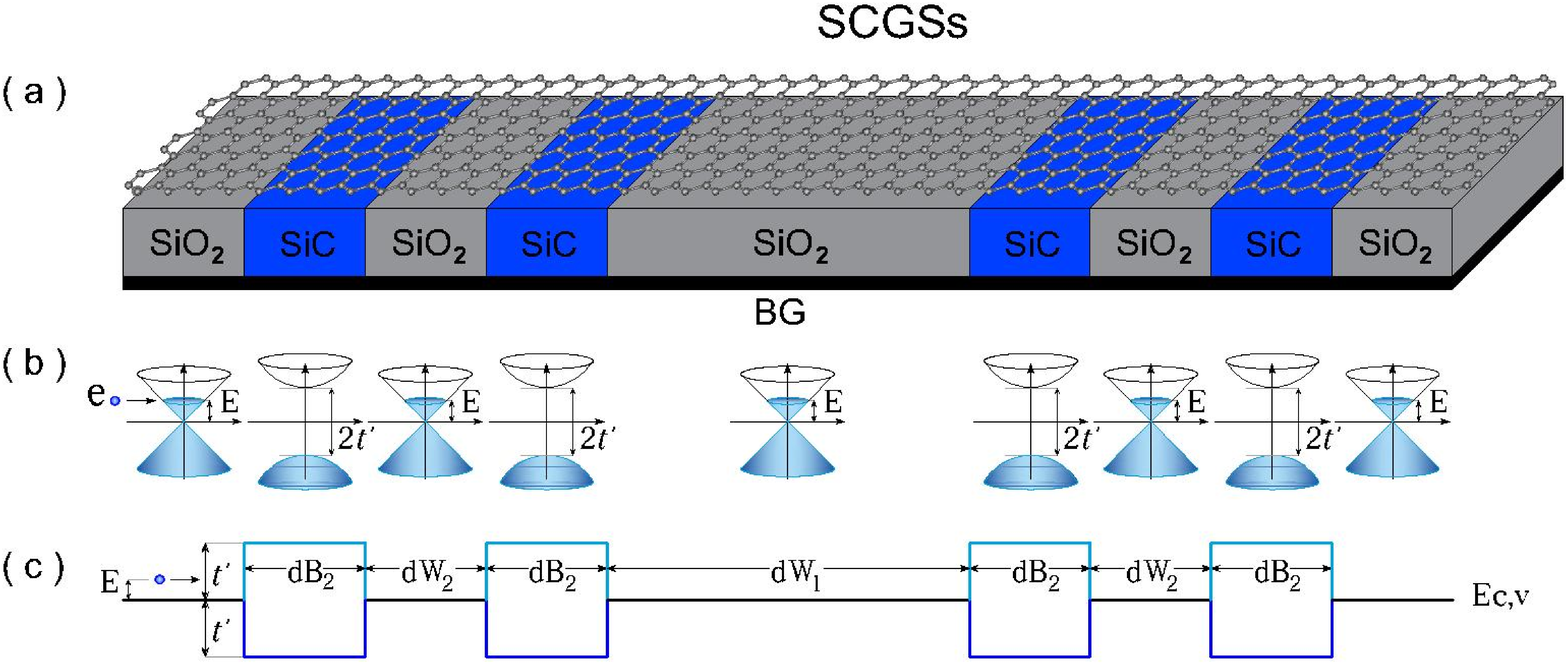}
\caption{\label{SCGSs} (Color online) Graphical representation of Substrate-Based Cantor Graphene Structures (SCGSs). (a) Representation of the cross-section of SCGSs. This structure is generated by the interaction between the substrate and the graphene layer. The graphene sheet is deposited above alternating substrates, such as SiO$_2$ (gray slabs) and SiC (blue slabs), obtaining regions with linear and parabolic dispersion relations, respectively. (b) Distribution of the Dirac cones and Dirac paraboloids along the structure. The bandgap in the Dirac paraboloids is $E_g=2t'$, the blue filled area indicates occupied states and the blue sphere describes an impinging massive Dirac electron at the Fermi energy. (c) Band-edge profile of SCGSs, the barriers have a height $V_0=t'$, and the widths of barriers and wells are dB$_{2}=w/9$, dW$_{2}=w/9$ and dW$_{1}=w/3$. The cyan and blue lines correspond to electrons and holes, respectively. All the illustrations are presented for the generation $N=3$.}
\end{figure}

\begin{figure}[h]
\includegraphics[scale=0.28]{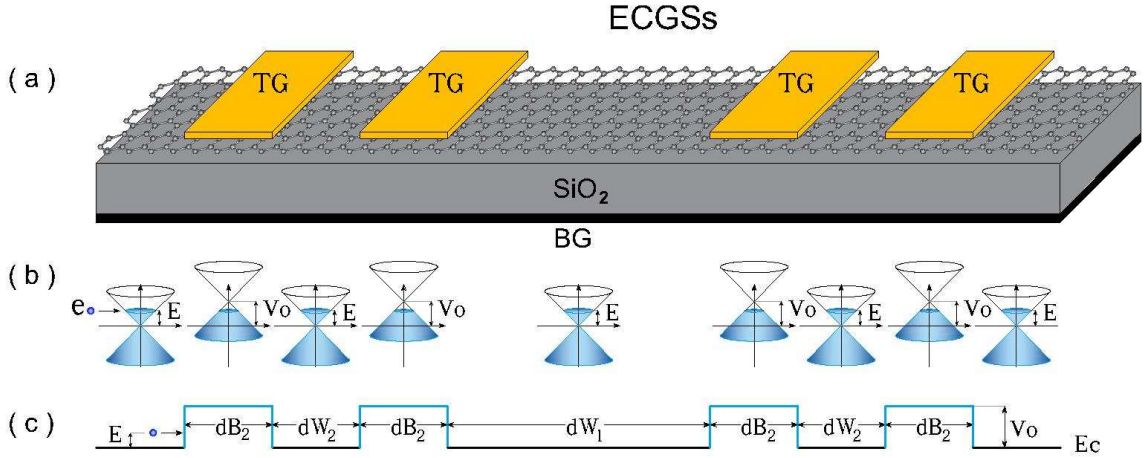}
\caption{\label{ECGSs} (Color online) Graphical representation of Electrostatic Cantor Graphene Structures (ECGSs). (a) Representation of the cross-section of ECGSs. This structure is generated when a graphene sheet is sited on a non-interacting substrate such as SiO$_2$ (gray slab) along with a back gate (black slab) and alternating top gates (yellow slabs). (b) Dirac-cone distribution along the structure. In this case, V$_{0}$ is proportional to the strength of the voltage applied to the back and top gates. The blue sphere describes an impinging massless Dirac electron at the Fermi energy. (c) Band-edge profile of ECGSs with  V$_{0}$ as height of the barriers. The cyan line corresponds to electrons, since we are considering positive voltages. All the illustrations are presented for the generation number $N=3$.}
\end{figure}

\begin{figure}[h]
\includegraphics[scale=0.28]{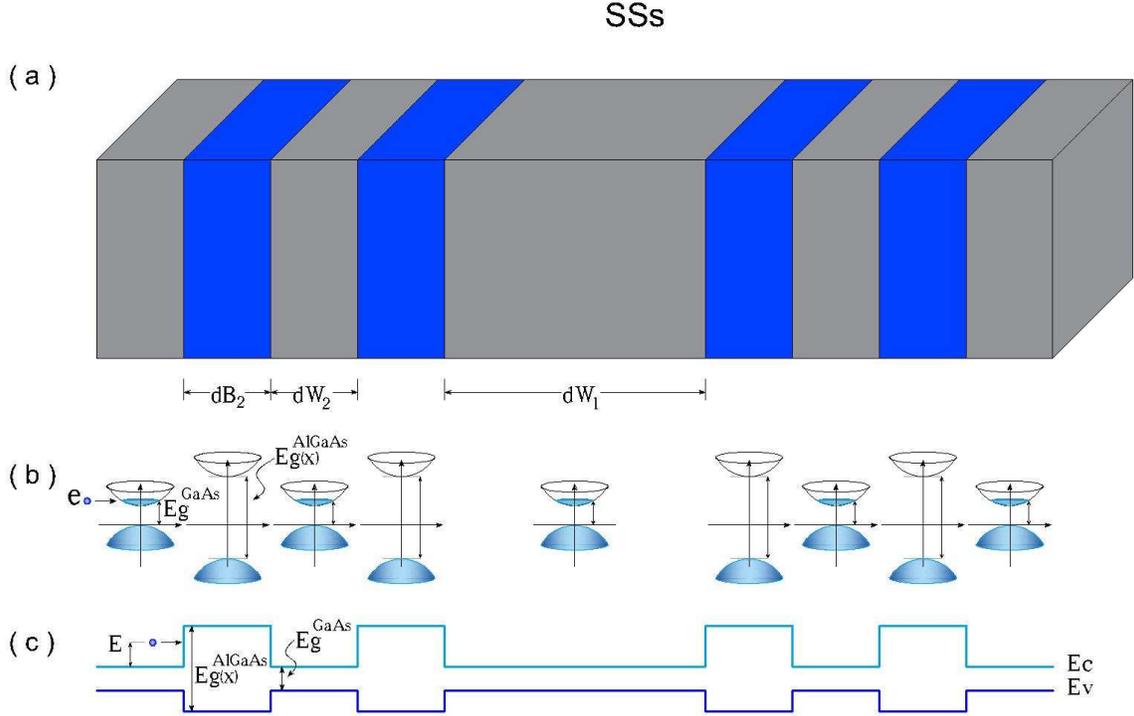}
\caption{\label{SSs} (Color online) Graphical representation of typical Al$_{x}$Ga$_{1-x}$As/GaAs Semiconductor Structures (SSs). (a) Representation of the cross-section of SSs. This structure can be created by alternating semiconductors with dissimilar bandgaps, such as GaAs (gray slabs) and AlGaAs (blue slabs). (b) Dirac-paraboloid distribution, the blue filled area indicates occupied states and the blue sphere describes an impinging Schr\"{o}dinger electron at the Fermi energy. (c) Band-edge profile of SSs. The cyan and blue lines correspond to electrons and holes, respectively. The height of the barriers is modulated by the aluminum concentration ($x$). All the illustrations are presented for the generation number $N=3$.}
\end{figure}

Prior to proceeding with the mathematical apparatus that we employed, we will present a simple procedure, an iterative method, to obtain the triadic Cantor set sequence. For the first generation, $N=1$, we choose a segment of any length. For the second generation, $N=2$,  we divide it into three equal parts and then eliminate the central one.  Finally, if we repeat this procedure at the $N-th$ generation the triadic Cantor set is obtained. Therefore, with the aforesaid in the previous section it is possible to develop a fractal arrangement of barriers according to the substitution rules of this aperiodic sequence, see Fig. \ref{SCGSs}, \ref{ECGSs} and \ref{SSs}. As our main objective is fractal structures in graphene, we will present the fundamentals of those structures in which the potential barriers are generated by substrates (Fig. \ref{SCGSs}) and electrostatic fields (Fig. \ref{ECGSs}). The fundamentals of Schr\"odinger electrons can be found elsewhere,\cite{DSDan2008PERL} and some minor details can be seen Fig. \ref{SSs}.   

To this regard, in graphene substrates such as silicon carbide (SiC) play an important role, since they are capable to change the form of the dispersion relation, from linear to parabolic,\cite{JVGomes2008JPCM} as well as to open a bandgap in the energy spectrum. \cite{SZhou2007NMaterials} The Hamiltonian that describes this kind of systems is given by, 

\begin{equation}
H=v_F (\mbox{\boldmath $\sigma\cdot p$}) + t'\sigma_z ,
\end{equation}

\noindent where $v_F$ is the Fermi velocity of quasi-particles in graphene of the order of $c/300$, $t' = mv_{F}^{2}$ is the mass term, $\mbox{\boldmath $\sigma$}=(\sigma_x,\sigma_y)$ are the Pauli matrices, $\textbf{\textit{p}}=(p_x,p_y)$ is the momentum operator and $\sigma_z$ the $z$-component of the Pauli matrix. This equation can be solved straightforwardly giving the following dispersion relation

\begin{equation}
E=\sqrt{\hbar^{2}v^{2}_{F}q^{2}+t'^{2}},
\label{parabolic_relation}
\end{equation}   

\noindent here $q$ is the wave vector associated with the SiC region, and $t'$ is proportional to the bandgap, $E_g=2t'$. Henceforth $t'$=0.13 eV, because we deal with a monolayer graphene system. In the case of the wavefunction we have,\cite{JVGomes2008JPCM} 

\begin{equation}
\psi_{\pm}(x,y) = \frac{1}{\sqrt{2}}
\left(\begin{array}{c}
1 \\
v_{\pm}
\end{array}\right)
e^{\pm iq_{x}x + iq_{y}y},
\label{wavefunction_SCGSs}
\end{equation} 		
\noindent with
\begin{equation}
v_{\pm} = \frac{\hbar v_{F}(\pm q_{x}+iq_{y})}{E+t'},
\end{equation}      

\noindent where the two components of the wavefunction are referred for the two equivalent triangular sublattices A and B in the graphene hexagonal structure, for this reason the wavefunction is considered as a bispinor.

On the other hand, systems with electrostatic barriers can be described by the following Hamiltonian, 

\begin{equation}
H=v_F(\mbox{\boldmath $\sigma\cdot p$}) + V(x),
\end{equation}

\noindent with $V(x)=V_0$ the one-dimensional potential along the $x$-direction. Considering the above Hamiltonian, the Dirac equation can be solved directly giving the linear dispersion relation, 

\begin{equation}
E=\pm\hbar v_{F}q + V_0,
\label{linear_relation}
\end{equation}   

\noindent where ``$\pm$'' represent electrons and holes, respectively. The wavefunctions are represented with the same mathematical form that in the case of substrate barriers, eq. (\ref{wavefunction_SCGSs}), but now the bispinor coefficients are written as, 

\begin{equation}
v_{\pm} = \frac{\hbar v_{F}(\pm q_{x}+iq_{y})}{E-V_0}.
\end{equation}         

Once the explicit form of the dispersion relations and wavefunctions are obtained for both substrate and electrostatic structures, it is easy to compute the transmission properties of our aperiodic fractal system by means of the transfer matrix formalism.\cite{PYeh2005, CSoukoulis2008} Specifically, the transmission probability or transmittance comes as, 

\begin{equation}
T=\left | \frac{A_{N+1}}{A_0} \right |^{2} = \frac{1}{| M_{11} |^2},
\end{equation}

\noindent where the transmitted wave amplitude $A_{N+1}$ can be calculated in terms of the incident wave amplitude $A_{0}$ via the transfer matrix,    

\begin{equation}
\left(\begin{array}{c}
A_0 \\
B_0
\end{array}\right)
=M
\left(\begin{array}{c}
A_{N+1} \\
0
\end{array}\right),
\end{equation} 

\noindent where the transfer matrix is given by,\cite{IRVargas2012JAP, JBTorres2014SM, RRGonzalez2015PE} 

\begin{equation}
M=D_{0}^{-1}\left[ \prod_{j=1}^{N}D_{j}P_{j}D_{j}^{-1}\right]D_{0}.
\end{equation} 

\noindent with $D_j$ and $P_j$ the dynamical and propagations matrices of the structures.\cite{IRVargas2012JAP, JBTorres2014SM, RRGonzalez2015PE}

\section{Results and discussion}
First of all, it is important to emphasize that the aim of our work is to find out the scaling properties in the transmission spectra for three fundamental parameters: generation $N$, height of the barriers $V_{0}$ and length of the system $w$. So, we study the self-similarity and scalability of the transmission spectra of SCGSs (Fig. \ref{SCGSs}) , ECGSs (Fig. \ref{ECGSs}) and a typical SSs (Fig. \ref{SSs}). Hereafter, we will focus our attention in three special mathematical expressions that can describe the scalability of our systems. 

Before continuing with the presentation of our numerical results, we present important aspects of the systems of interest: SCGSs, ECGSs and SSs. In the case of SCGSs, the charge carriers are treated as relativistic particles known as massive Dirac electrons. Other characteristic is that the Klein effect is ruled out due to the non-conservation of the pseudo-spin. In the case of ECGSs, the charge carriers are massless Dirac electrons. In this case, at normal incidence the Klein effect, perfect transmission, is manifested. At this point, it is important to mention that our results for massless Dirac electrons at normal incidence are not illustrative, in view of the fact that the transmittance is always $T=1$, irrespective of the generation, height and length of the system, and the scaling rules will be always valid. So, the results for massless Dirac electrons at normal incidence will be not presented.  In the case of SSs, the charge carriers are Schr\"{o}dinger electrons. An important characteristic in this case is that the transversal motion of Schr\"{o}dinger electrons is separable, so the transmission properties are not dependent of the angle of incidence.  Therefore, we will only present results for Schr\"{o}dinger electrons at normal incidence.

In order to analyse the scaling rules of the transmission spectra between generations, we propose an expression that depends on the factor three, which is involved when one changes the generation in the triadic Cantor set. Specifically, the expression comes as,

\begin{equation}
T_{N}(E)\approx [T_{N+1}(E)]^{3}.
\label{generation}
\end{equation}

\begin{figure}[h] 
\includegraphics[scale=0.6]{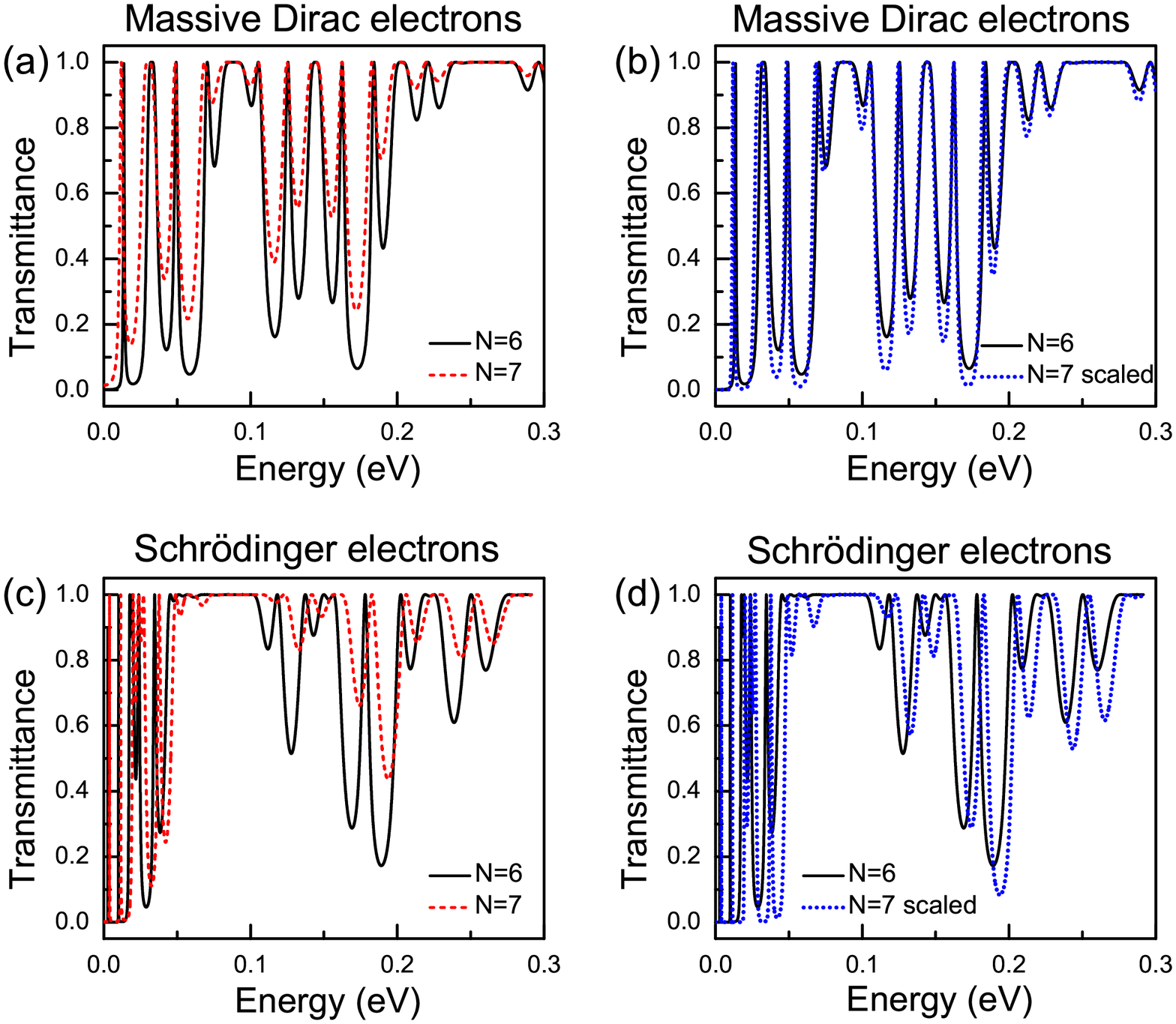}
\caption{\label{ScalingGenerationsNormal} (Color online) Scaling between generations at normal incidence for SCGSs (first row) and SSs (second row). (a) Comparison of the transmittance as function of energy for generations $N=6$ (solid-black lines) and $N=7$ (dashed-red lines). (b) Using the appropriate scaling equation (eq. \ref{generation}), we can corroborate that the reference ($N=6$) and scaled ($N=7$) curves are approximately the same, solid-black and dotted-blue lines, respectively. (c) and (d) correspond to the same comparison as in (a) and (b) but for SSs, or equivalently Schr\"odinger electrons. In this case the scalability is not present. The structural parameters of both systems SCGSs and SSs are $w=1000a$ and $V_{0}=0.13$ eV.} 
\end{figure}

Now, we implement this expression for massive Dirac electrons and Schr\"{o}dinger electrons at normal incidence, see Fig. \ref{ScalingGenerationsNormal}. The structural parameters for both systems are the same, that is, the height and width of the barriers as well as the generation are the same. In particular, we consider barriers of height and starting width of $V_0=0.13$ eV and $w=1000a$, with $a=1.42$ \AA \ the carbon-carbon distance in graphene, and generations $N=6$ and $N=7$. From now on, $N=6$ will be the generation of reference and $N=7$ the generation to be scaled or simply the scaled generation. In Fig. \ref{ScalingGenerationsNormal}(a) we show the transmittance as a function of the energy for massive Dirac electrons. The solid-black and the dashed-red curves correspond to generations 6 and 7, respectively. As we can see the envelopes of these curves are quite similar, the only difference that we can notice is a shift in the vertical axis. Appealing to this similarity we implement our scale rule (eq. \ref{generation}) to obtain the transmission spectrum of generation $N=6$ by means of the scaling of generation $N=7$, that is, $T_{6}(E) \approx [T_{7}(E)]^{3}$. As we can see in Fig. \ref{ScalingGenerationsNormal}(b) the matching between the generation of reference ($N=6$) and the scaled one ($N=7$) is quite good, compare the solid-black and dotted-blue curves. In the case of 
Schr\"{o}dinger electrons the transmission spectra between generations 6 and 7 are also similar, see the solid-black and dashed-red curves in Fig. \ref{ScalingGenerationsNormal}(c). However, in this case, in addition to the shift in the vertical axis, we can see an offset between the curves in the energy axis. And it is precisely this difference that makes that our scale rule not work well in the case of 
Schr\"{o}dinger electrons, compare the solid-black and dotted-blue curves in Fig. \ref{ScalingGenerationsNormal}(d). 

\begin{figure}[h]
\includegraphics[scale=0.6]{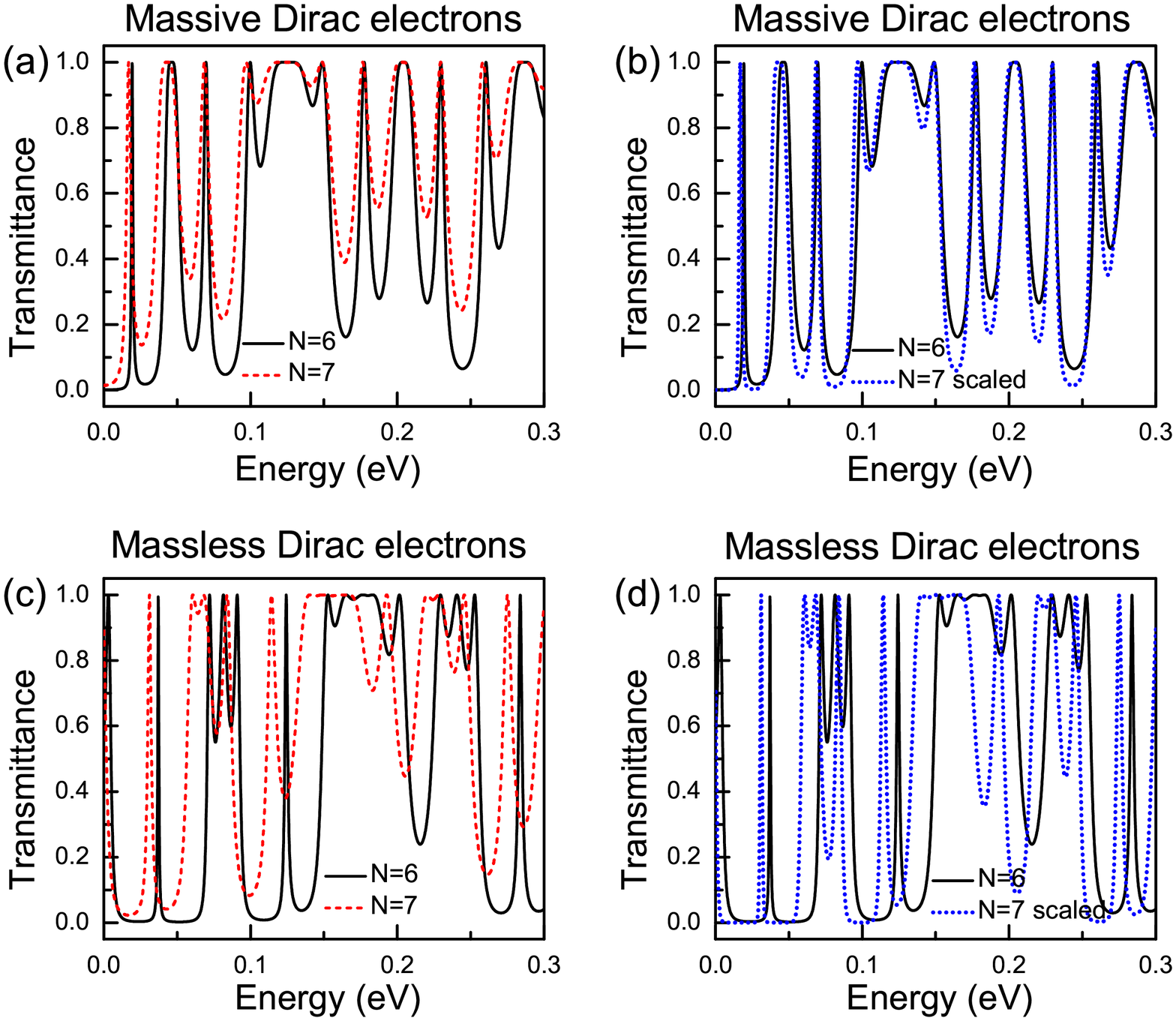}
\caption{\label{ScalingGenerationsOblique} (Color online) Scaling between generations at oblique incidence ($\theta=\pi/4$) for SCGSs (first row) and ECGSs (second row). (a) Comparison of the transmittance as function of energy for generations $N=6$ (solid-black lines) and $N=7$ (dashed-red lines). (b) Using the appropriate scaling equation (eq. \ref{generation}), we can corroborate that the reference ($N=6$) and scaled ($N=7$) curves are approximately the same, solid-black and dotted-blue lines, respectively. (c) and (d) correspond to the same comparison as in (a) and (b) but for ECGSs, or equivalently massless Dirac electrons. In this case the scaling properties are not visible. The structural parameters are the same as in Fig. \ref{ScalingGenerationsNormal}.}
\end{figure}

The scaling properties of massive and massless Dirac electrons at oblique incidence ($\theta=\pi/4$) are presented in Fig. \ref{ScalingGenerationsOblique}. These cases are in contraposition, since in one case the conservation of the pseudo-spin is ruled out (massive Dirac electrons), while in the other case it is sustained (massless Dirac electrons). The structural parameters are the same as in Fig. \ref{ScalingGenerationsNormal}. For massive Dirac electrons, once again, we can see that the transmission spectra between generations 6 and 7 are quite similar, the main difference is a relative shift in the transmission axis, see Fig. \ref{ScalingGenerationsOblique}(a). By applying our scale rule to $T_7(E)$, we obtain a very good correspondence with respect to the transmission spectrum of reference ($T_6(E)$), compare the solid-black and dotted-blue curves in Fig. \ref{ScalingGenerationsOblique}(b). The main differences between these curves are localized in the minimums of the transmittance. In the case of massless Dirac electrons, we can notice that the scaling properties are worsen in comparison with the massive case, see Fig. \ref{ScalingGenerationsOblique}(d).  Even, this case is worse with respect to the the case of Schr\"{o}dinger electrons, compare Figs. \ref{ScalingGenerationsNormal}(d) and \ref{ScalingGenerationsOblique}(d). These significant differences between the reference spectrum and the scaled one come from the large shifts in both the transmittance and energy axes between $T_6(E)$ and $T_7(E)$, see Fig. \ref{ScalingGenerationsOblique}(c).   
   
Continuing with our study about CGSs, we have also found that self-similar patterns are visible when two systems that have a different height of the barriers are considered. Therefore, the scalability for SCGSs, ECGSs and SSs is examined. Thus, comes the turn to view how the heights of the barriers play a fundamental role with respect to the self-similar properties of our aperiodic structures. In this regard, the following expression is proposed, in general, to analyse the scaling properties between systems with different heights of the barriers or simply the scaling between heights,\cite{DSDan2015} 

\begin{equation}
T_{V_{0}}(E)\approx[T_{\frac{1}{k}V_{0}}(E)]^{k^{2}},
\label{height}
\end{equation} 

\noindent where the subindex $V_{0}$ indicates the height of the barriers, and $k$ is a factor that comes in multiples of two.

\begin{figure}[h]
\includegraphics[scale=0.6]{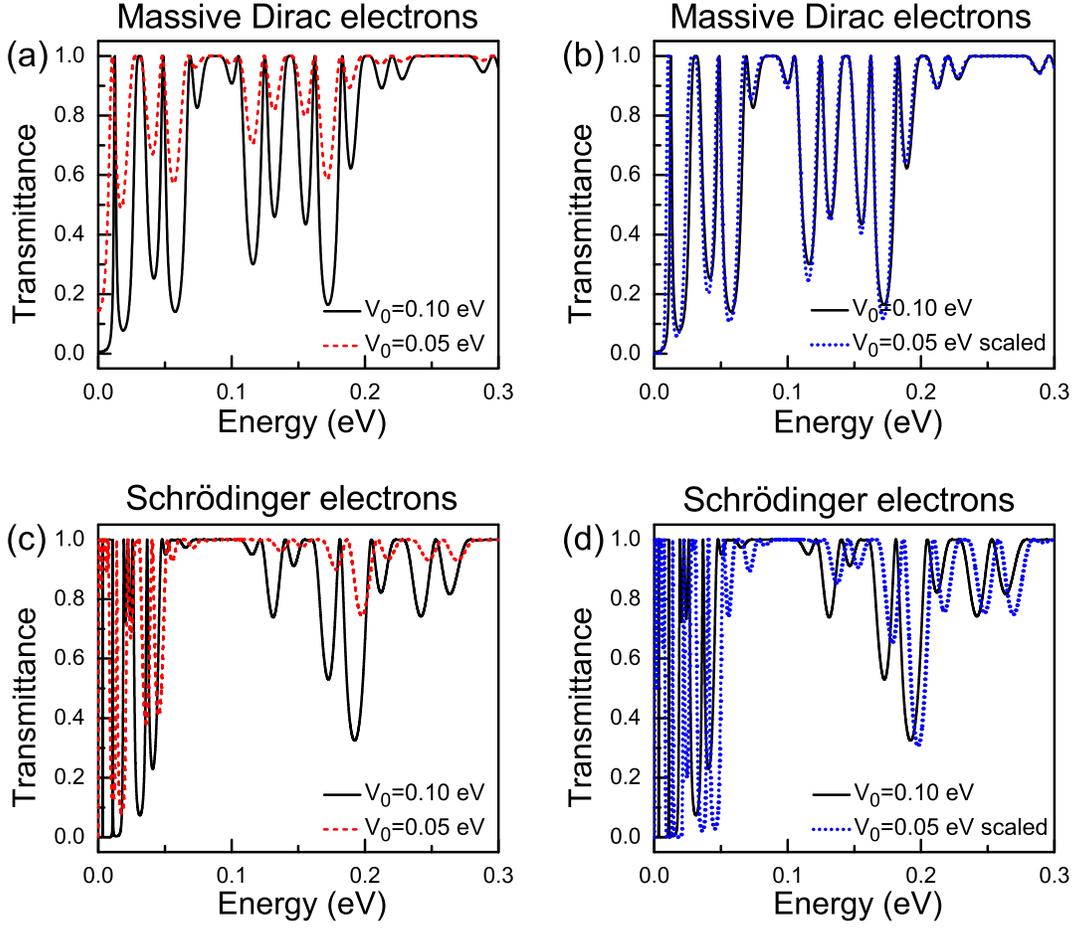}
\caption{\label{ScalingHeightsNormal} (Color online) Scaling between heights of the barriers at normal incidence for SCGSs (first row) and SSs (second row). (a) Comparison of the transmittance as function of energy for the heights $V_{0}=0.10$ eV (solid-black lines) and $V_{0}=0.05$ eV (dashed-red lines). (b) By implementing the appropriate scaling equation (eq. \ref{height}), we can see that the scaled curve corresponding to $V_{0}=0.05$ eV is practically similar to the reference one $V_{0}=0.10$ eV, dotted-blue and solid-black lines, respectively. (c) and (d) correspond to the same comparison as in (a) and (b) but for SSs. In this case the scalability is not present. The structural parameters for both systems, SCGSs and SSs, are $w=1000a$ and $N=6$.}
\end{figure}

To this respect, Fig. \ref{ScalingHeightsNormal} depicts the scaling between heights at normal incidence ($\theta=0$) for both massive Dirac electrons and  Schr\"{o}dinger electrons. In this case the structural parameters are $N=6$ and $w=1000a$. At first, we show comparatively  the transmission spectra for a structure with two different heights of the barriers. In particular, Figs. \ref{ScalingHeightsNormal}(a) and (c) correspond to massive Dirac electrons and Schr\"{o}dinger electrons, respectively. Specifically, the transmission spectrum is computed for a height equal to $V_{0}=0.10$ eV and $V_{0}=0.05$ eV. Both spectra denoted by $T_{0.10}(E)$ and $T_{0.05}(E)$, solid-black and dashed-red curves, respectively. Then, if we implement the scale rule (eq. \ref{height}) for these heights, that is, $T_{0.10}(E)\approx [T_{0.05}(E)]^{4}$, a scaled transmission spectrum for SCGSs (Fig. \ref{ScalingHeightsNormal}(b)) and for SSs (Fig. \ref{ScalingHeightsNormal}(d)) is obtained. If we paid attention of the results for massive Dirac electrons we can see a great matching between the reference transmission and the scaled transmission. Thus, we can corroborate that the self-similarity and scalability are exhibited in SCGSs, and even more interesting, we have found a scaling rule that describe these properties. On the other hand, Fig. \ref{ScalingHeightsNormal}(d) displays the scaling between heights for the case of Schr\"{o}dinger electrons. The height of the barriers for this system, Al$_{x}$Ga$_{1-x}$As/GaAs, is modulated changing the aluminum concentration ($x$). Here, we can see that the reference transmission and the scaled transmission, solid-black and dotted-blue curves, have dissimilar envelopes, that is, the scale rule is not fulfill at all. Even more, this dissimilarity is easy to visualize in the mentioned figure because the reference transmission and the scaled transmission are shifted one respect to other in both axes.

\begin{figure}[h]
\includegraphics[scale=0.6]{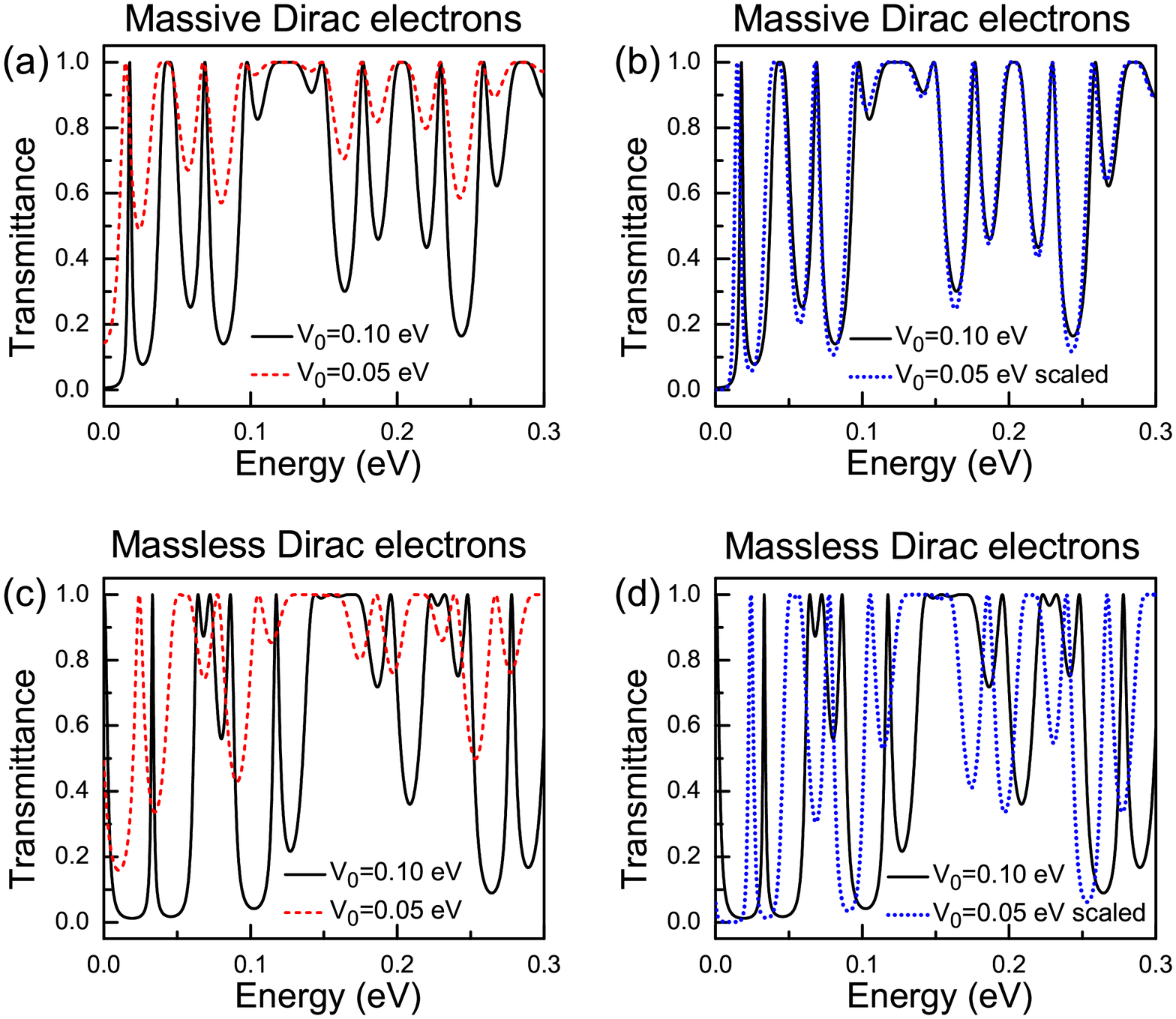}
\caption{\label{ScalingHeightsOblique} (Color online) The same as in Fig. \ref{ScalingHeightsNormal}, but for SCGSs (first row) and ECGSs (second row), at oblique incidence $\theta=\pi/4$. As in Fig.  \ref{ScalingHeightsNormal} massive Dirac electrons present excellent scaling properties.}
\end{figure}

In Fig. \ref{ScalingHeightsOblique} we show the scaling properties between heights of the barriers, at oblique incidence ($\theta=\pi/4$), for massive and massless Dirac electrons. All the parameters for SCGSs and ECGSs are the same as in Fig. \ref{ScalingHeightsNormal}. The corresponding transmission spectrum for massive and massless Dirac electrons is calculated for two different heights of the barrier, Figs. \ref{ScalingHeightsOblique}(a) and (c), respectively. At first sight, the transmission spectra for massive Dirac electrons seem different, however, they have something in common, that is, both curves show a similar envelope, compare the solid-black and dashed-red curves in Fig. \ref{ScalingHeightsOblique}(a). So, it is natural to think in a scale rule that reproduces one curve with respect to the other. By applying eq. \ref{height} to the transmittance of  height $V_{0}=0.05$ eV, we obtain the corresponding scaled curve, dotted-blue curve in Fig. \ref{ScalingHeightsOblique}(b). As we can corroborate, a very good matching between the reference and scaled transmittance is obtained. The main differences are found in the minimums of the transmission. Even, it seems that the matching is better than in the case of the scaling between generations, see Fig. \ref{ScalingGenerationsOblique}(b). In contrast, Fig. \ref{ScalingHeightsOblique}(d) clearly shows that this scale rule is not suitable for massless Dirac electrons. We can notice a remarkable offset between the curves, in both the transmission and energy axis. Even more, the envelopes of the curves are not corresponding, see Figs. \ref{ScalingHeightsOblique}(c) and (d). In short, we can say that the scaling between heights is valid for massive Dirac electrons, irrespective if the electrons impinge at normal or oblique incidence.  
 
Another self-similar patterns in the transmission properties occur when a structure with two different lengths is taken into account. With the length of the system we mean to the width of the barrier in the first generation of the structure. Within this context, the following expression is used to study this kind scaling property,\cite{DSDan2015}

\begin{equation}
T_{w}(E) \approx [T_{\frac{1}{\alpha}w}(\frac{1}{\alpha}E)]^{\alpha^{2}},
\label{length}
\end{equation}

\noindent where the subindex $w$ refers to the length of the system, and the factor $\alpha$ comes in multiples of two. At this point, it is important to remark that this expression is a little bit different with respect to the former ones, since in addition to scale the transmission, it is required that the argument (the energy) of it, be scaled as well. 

\begin{figure}[h]
\includegraphics[scale=0.6]{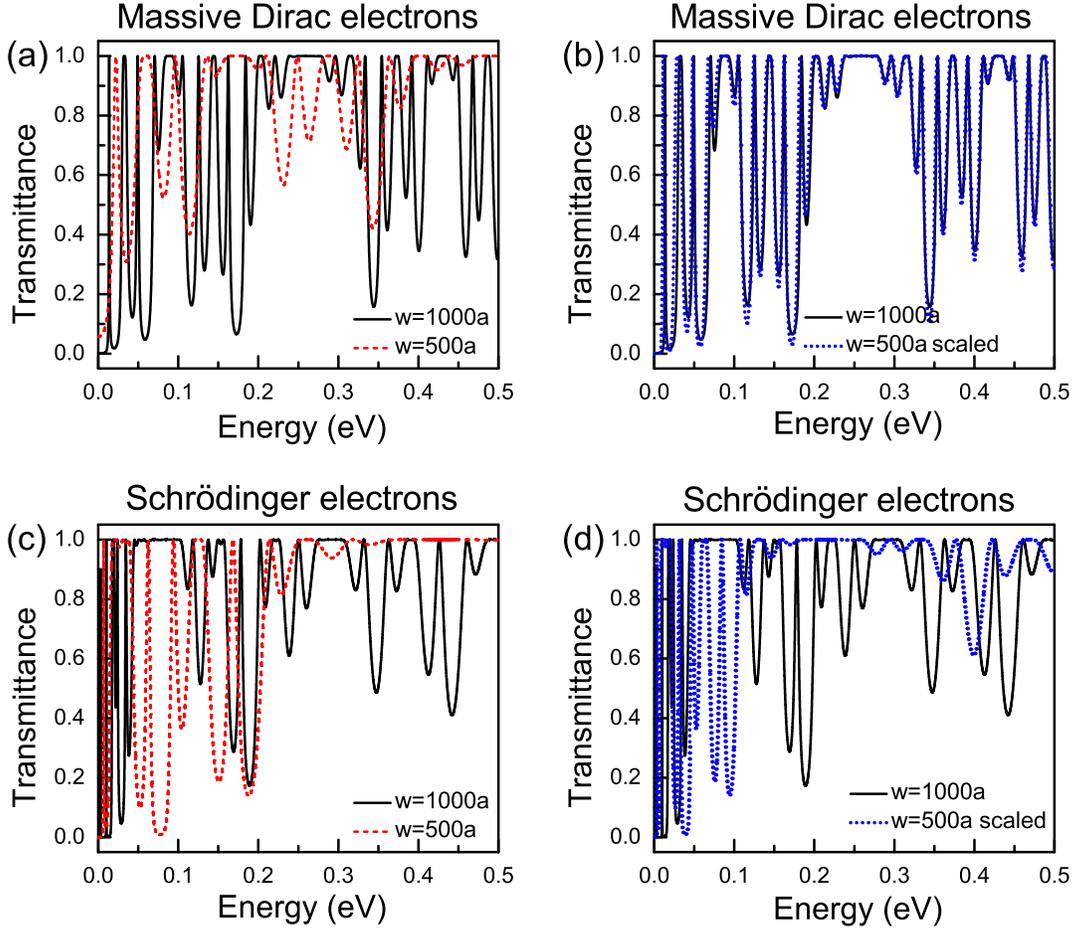}
\caption{\label{ScalingLengthsNormal} (Color online) Scaling between lengths of the system at normal incidence for SCGSs (first row) and SSs (second row). (a) Comparison of the transmittance as function of energy for lengths $w=1000a$ (solid-black lines) and $w=500a$ (dashed-red lines). (b) By applying eq. \ref{length} the scaled curve ($w=500a$) is obtained, which has an excellent match with the reference one ($w=1000a$). (c) and (d) correspond to the same comparison as in (a) and (b) but for SSs. In this case the scalability is not present. Here, the structural parameters are $N=6$ and $V_0=0.13$ eV.}
\end{figure}

\begin{figure}[h]
\includegraphics[scale=0.6]{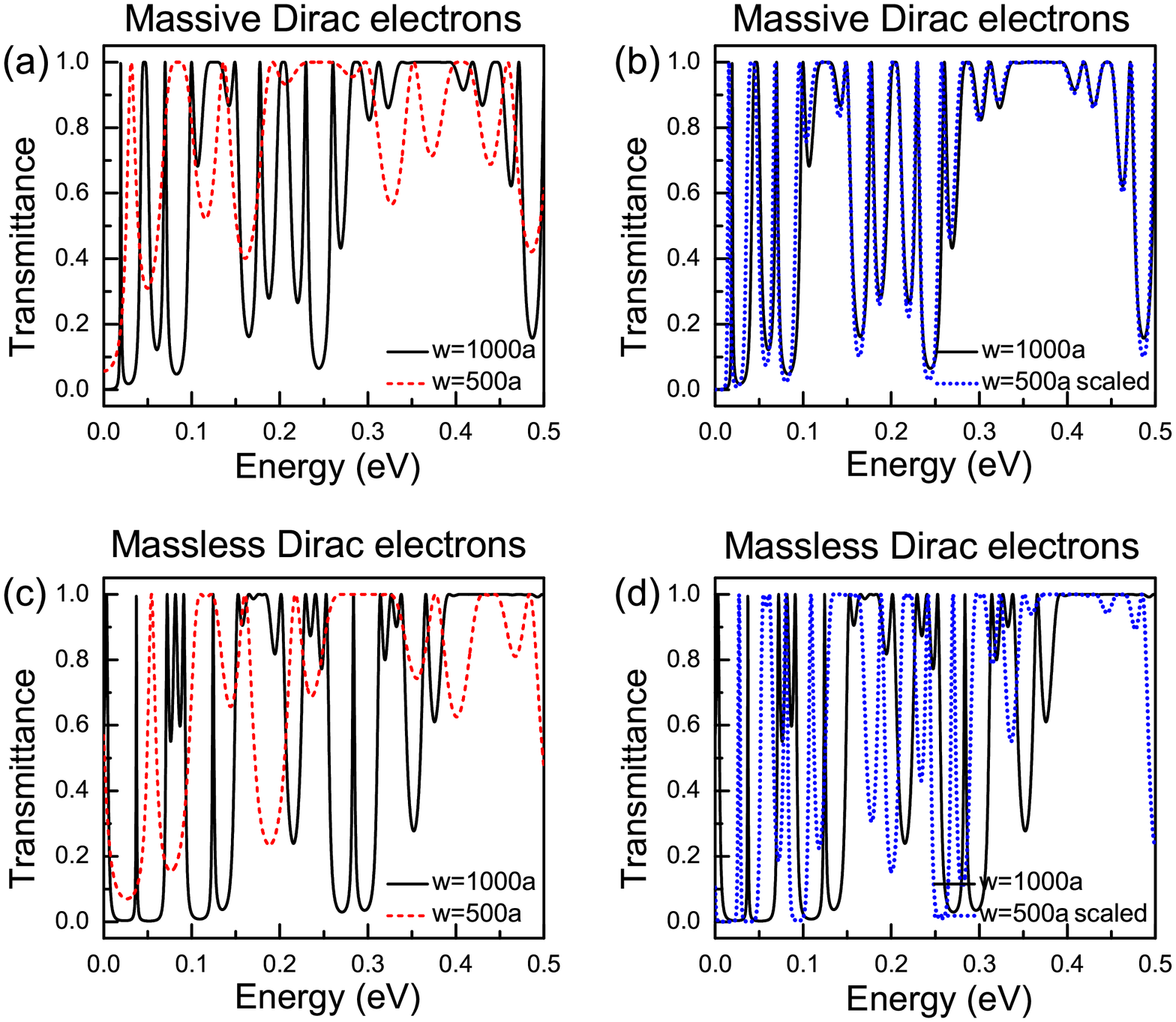}
\caption{\label{ScalingLengthsOblique} (Color online) The same as in Fig. \ref{ScalingLengthsNormal},
but for SCGSs (first row) and ECGSs (second row), at oblique incidence $\theta=\pi/4$. As in Fig. \ref{ScalingLengthsNormal} massive Dirac electrons present excellent scaling properties.}
\end{figure}

By applying this expression to the case of massive Dirac electrons and Schr\"{o}dinger electrons at normal incidence (Fig. \ref{ScalingLengthsNormal}), we can see that a very good scaling is obtained for massive Dirac electrons, see Fig. \ref{ScalingLengthsNormal}(b). In this case the structural parameters are $N=6$ and $V_{0}=0.13$ eV. We have chosen as lengths of the system $w=1000a$ and $w=500a$, the former representing the reference system and the latter the scaled one. In addition, the factor $\alpha$ is equal to 2, so the reference and scaled transmissions are related by $T_{1000a}(E)\approx [T_{500a}(\frac{1}{2}E)]^{4}$. The matching between these curves is quite surprising, since at first sight they are totally dissimilar, compare solid-black and dashed-red curves in Fig. \ref{ScalingLengthsNormal}(a). Now considering the case for Schr\"{o}dinger electrons, see Fig. \ref{ScalingLengthsNormal}(d), turns out that the reference and scaled transmission are practically different. Even, this case is worsen as compare with the other scale rules, see Figs. \ref{ScalingGenerationsNormal}(d) and \ref{ScalingHeightsNormal}(d). So, as far as we can see the transmission properties of Schr\"{o}dinger electrons do not manifest a self-similar behaviour.
On the other hand, if we apply this scale rule to the propagation of electrons in SCGSs and ECGSs, we find that a very good scaling is presented in the case of massive Dirac electrons, see Fig. \ref{ScalingLengthsOblique}. Even, when in this case the scaling for massless electrons is better than the corresponding one for Schr\"odinger electrons, see Figs. \ref{ScalingLengthsNormal}(d) and \ref{ScalingLengthsOblique}(d), the difference between the reference and the scaled curves is significant, compare the solid-black and the dotted-blue curves in Fig. \ref{ScalingLengthsOblique}(d).

\begin{figure}[h]
\includegraphics[scale=0.6]{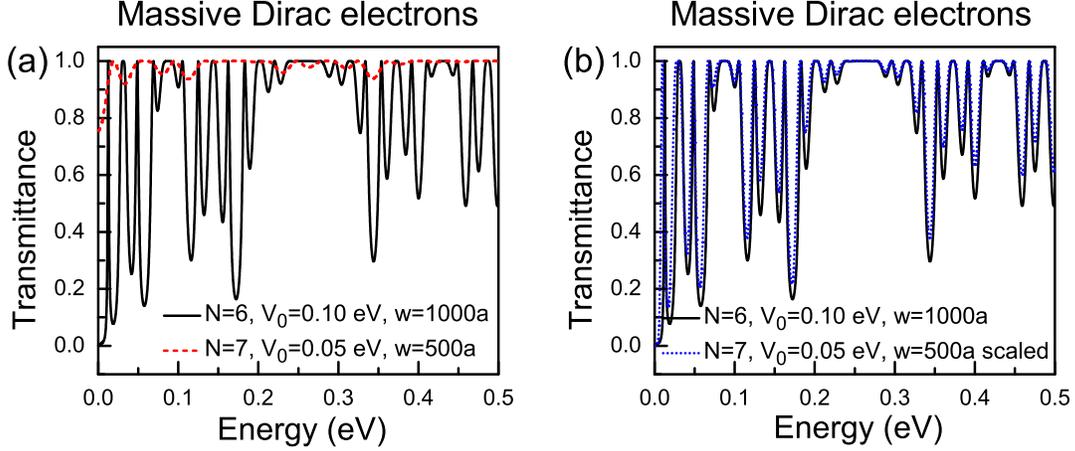}
\caption{\label{GeneralScaling} (Color online) General scaling at normal incidence for SCGSs. (a) Comparison of the transmittance as function of energy for $N=6$, $V_{0}=0.10$ eV and $w=1000a$ (solid-black lines) and $N=7$, $V_{0}=0.05$ eV and $w=500a$ (dashed-red lines). (b) By applying eq. \ref{general-scaling} the scaled curve (dotted-blue lines) is obtained, which has an excellent match with the reference one (solid-black lines). As in all previous figures massive Dirac electrons present outstanding scaling properties.}
\end{figure}

Even more interesting, the three scaling rules can be unified by means of the following general scaling expression:

\begin{equation}
T(E, N, V_{0}, w)\approx [T(\frac{1}{\alpha}E, N+m, \frac{1}{k}V_{0}, \frac{1}{\alpha}w)]^{3^{m}(k\alpha)^{2}},
\label{general-scaling}
\end{equation}

\noindent where the factors \textit{k} and $\alpha$ come in multiples of two, and \textit{m} is the difference between generations. As we can see in Fig. \ref{GeneralScaling}, this general rule works pretty well for massive Dirac electrons.

After analysing the different scaling rules, we obtain that, in general, the transmission properties of massive Dirac electrons present scalability, and as a consequence self-similarity. However, so far our comparison between the reference and scaled curves has been visual. What we need now is to get a quantitative measure of how similar are these curves. In order to address this issue, we compute the root mean square deviation or simply the rmsd, which is a standard measure to compare curves. If the curves tend to be exactly the same, this value is practically zero. For instance, the rmsd of massive Dirac electrons at normal incidence for the scaling between generations $N$, heights V$_{0}$ and lengths $w$ are 0.134303480, 0.110821329 and 0.133542433, respectively. In the same way, the corresponding ones for Schr\"{o}dinger electrons are: 0.256470233, 0.276356488 and 0.52796197. In the case of massive Dirac electrons at oblique incidence these values are 0.159395278, 0.131747648 and 0.157041878, while the corresponding ones for massless Dirac electrons become 0.448230058, 0.478004634 and 0.463266015. Based on these results turns out that practically in all cases the rmsd is smaller for massive Dirac electrons. In particular, the rmsd for massive electrons is 3.6 times smaller than the rmsd of massless electrons in the case of the scaling between heights. This value becomes 2.5, if we compare massive Dirac electrons and Schr\"odinger electrons. At this point, it is also important to emphasize that if we change the energy range up to 1 eV, the matching between the reference and scaled transmission for massive Dirac electrons is even better. Specifically, the rmsd becomes one order of magnitude smaller. As a final remark, we can say that from both the qualitative and quantitative standpoints, the transmission properties of massive Dirac electrons are self-similar. 

\section{Conclusions}
In summary, we have studied the self-similarity and scalability of the transmission properties of Cantor Graphene Structures. The transfer matrix formalism has been used to compute the transmission spectra. In particular, three mathematical expressions are proposed and implemented to this end, finding three different scaling rules, which are associate with three fundamental parameters: the generation number, the height of the barriers and the length of the system. Thus, a comparative analysis of the scaling between generations, heights and lengths of the system for massive and massless Dirac electrons as well as Schr\"{o}dinger electrons is carried out, obtaining that the better scaling is for massive Dirac electrons. This result is valid for both normal and oblique incidence. Finally, as far as we can see to obtain self-similar characteristics in the physical properties, such as the transmittance, are necessary three conditions: relativistic Dirac electrons, a self-similar structure and the non-conservation of the pseudo-spin (massive Dirac electrons).

\acknowledgments
R. R. -G. acknowledges to CONACYT-Mexico for the scholarship for doctoral studies.

\pagebreak


\end{document}